# Angle-Resolved Magneto-Chiral Anisotropy in a Non-Centrosymmetric Atomic Layer Superlattice

*Long Cheng,[1,†] Mingrui Bao,[1,†] Jingxian Zhang,[1,2,†] Xue Zhang,[3,†] Qun Yang,[1] Qiang Li,[4] Hui Cao,[5] Dawei Qiu,[2] Jia Liu,[1] Fei Ye,[1] Qing Wang,[1] Genhao Liang,[1,2] Hui Li,[6] Guanglei Cheng,[2] Hua Zhou,[7] Jian-Min Zuo,[8] Xiaodong Zhou,[4] Jian Shen,[4] Zhifeng Zhu,[3*] Sai Mu,[9*] Wenbo Wang,[1,10*] Xiaofang Zhai[1*]*

[1]School of Physical Science and Technology, ShanghaiTech University, Pudong District, Shanghai 201210, China

[2]CAS Key Laboratory of Microscale Magnetic Resonance and School of Physical Sciences, University of Science and Technology of China, Hefei 230026, China

[3]School of Information Science and Technology, ShanghaiTech University, Pudong District, Shanghai 201210, China

[4]State Key Laboratory of Surface Physics and Institute for Nanoelectronic Devices and Quantum Computing, Fudan University, Shanghai 200433, China.

[5]Materials Science Division, Argonne National Laboratory, Lemont, IL 60439, USA

[6]Institutes of Physical Science and Information Technology, Anhui University, Hefei 230601, China

[7]X-ray Science Division, Advanced Photon Source, Argonne National Laboratory, Lemont, IL 60439, USA

[8]Department of Materials Science and Engineering and Materials Research Laboratory, University of Illinois Urbana–Champaign, Urbana, IL 61801, USA

[9]Center for Experimental Nanoscale Physics, Department of Physics and Astronomy, University of South Carolina, Columbia, South Carolina 29208, USA

[10]ShanghaiTech Laboratory for Topological Physics, ShanghaiTech University, Shanghai 201210, China

[†]These authors contributed equally to this work.
[*]Emails: zhaixf@shanghaitech.edu.cn; wangwb1@shanghaitech.edu.cn; mus@mailbox.sc.edu; zhuzhf@shanghaitech.edu.cn

**Abstract**

Chirality in solid-state materials has sparked significant interest due to potential applications of topologically-protected chiral states in next-generation information technology. The electrical magneto-chiral effect (eMChE), arising from relativistic spin-orbit interactions, shows great promise for developing chiral materials and devices for electronic integration. Here we demonstrate an angle-resolved eMChE in an A-B-C-C type atomic-layer superlattice lacking time and space inversion symmetry. We observe non-superimposable enantiomers of left-handed and right-handed tilted uniaxial magnetic anisotropy as the sample rotates under static fields, with the tilting angle reaching a striking 45°. Magnetic force microscopy and atomistic simulations correlate the tilt to the emergence and evolution of chiral spin textures. The Dzyaloshinskii-Moriya interaction 'lock' effect in competition with Zeeman effect is demonstrated to be responsible for the angle-resolved eMChE. Our findings open up a new horizon for engineering angle-resolved magneto-chiral anisotropy, shedding light on the development of novel angle-resolved sensing or writing techniques in chiral spintronics.

**Introduction**

Chirality is a widespread phenomenon in nature, where two non-superimposable enantiomers exist with one being the mirror image of the other. The interplay between chirality and magnetism has recently drawn intensive interests in non-centrosymmetric magnetic materials absent of both spatial and time inversion symmetries.[1-3] Among various chiral magnetic effects, the electrical manifestation of magneto-chiral effect is crucial for both fundamental science and electronic applications.[4-6] The eMChE has recently been intensively studied in topological semimetals, where the non-reciprocal resistance depending on the inner product of the magnetic field (**B**) and the current (**I**) is observed.[7-9] In non-centrosymmetric magnets, chiral magnetic interaction known as Dzyaloshinskii-Moriya interaction (DMI) exists due to the relativistic spin-orbit coupling (SOC) effect. Depending on the spin winding direction from site $i$ to site $j$, the sign of the DMI energy term $\boldsymbol{D}_{ij} \cdot (\boldsymbol{S}_i \times \boldsymbol{S}_j)$ manifests the chirality. DMI has been central to the magnetoelectric coupling in polar oxides[10-13]. However, most naturally existing magnets are centrosymmetric where DMI is absent, or being insulators, consequently making the eMChE a rare phenomenon. Therefore it is important to explore eMChE in artificially-structured chiral magnets where the centrosymmetry is deliberately removed, especially atomic layer-by-layer grown thin films where the inversion symmetry can be removed at the single atomic level.

In order to realize the eMChE, a chirality-dependent and electrically-measurable property of a material or a device is necessary. In analogous to DMI, the magnetic anisotropy (MA) also originates from SOC but with higher-order dependence[14,15], which is of central importance to magnetic device applications.[16] In centrosymmetric magnets, MA mainly originates from the magneto-crystalline anisotropy which determines the preferred and non-preferred spin orientations, i.e. the magnetic easy and hard axes. Different from DMI, MA is conveniently observable using electrical probes, such as the angular magnetoresistance (AMR).[17] On the other hand, in non-centrosymmetric magnet, DMI introduces chiral spin orders, such as spiral, helical, conical and skyrmion states.[3] The electrical manifestation of these chiral magnetic orders is very challenging. MA is known to significantly influence the formation, period and shape of chiral states. In turn, MA is affected by the formation, evolution and transition of these chiral states.[18] In non-centrosymmetric magnets, the contribution to MA is not limited to symmetric terms ($e.g.$ $|\nabla \boldsymbol{m}|^2$) from magneto-crystalline anisotropy, but also from anisotropic terms ($e.g.$ $|\nabla \times \boldsymbol{m}|^2$) from chiral exchange.[14] Therefore, MA is expected to depend on the chiral magnetic order. Indeed, neutron diffraction experiments on single-crystals have revealed magnetic easy axis reorientations in the conical phase regime of $Cu_2OSeO_3$ and the skyrmion lattice reorientation in MnSi.[14,18-20] Thus it is promising to find the chiral entity of MA, as a new type of eMChE, in artificially structured magnets absent of centrosymmetry.

Non-centrosymmetric heterostructures of $SrRuO_3$ (SRO) stands as a model system for probing eMChE since the Ru element is one of the heaviest transition metals that exhibit magnetism, thus providing a strong source of SOC.[21] Compressively strained SRO thin films exhibit perpendicular MA (PMA) which is an important property for spintronic applications.[22] The SRO heterostructure remains an active playground for engineering MA and fundamental studies of novel MA related phenomena[17,23-25]. Nevertheless, the chiral style of MA in non-centrosymmetric SRO heterostructures has not been studied.

Here we demonstrate an angle-resolved eMChE in an inversion-symmetry-broken A-B-C-C atomic layer superlattice, with A, B, C being single atomic layer of $SrRu_{1-x}Ti_xO_3$, SRO and SrTiO$_3$ (STO) respectively. The eMChE is demonstrated by AMR measurements, which show chiral tilts of the uniaxial magnetic easy axis depending on the sample rotation handedness. Magnetic force microscopy (MFM) experiments show characteristic transitions of chiral magnetism from stripe domains to magnetic bubbles, corresponding well to the development of angle-resolved MChE from a low tilting range of 2°-30° to a high tilting angle of 45°. Atomistic simulations reveal that the high tilting angle is related to the magnetic bubble formation. We explain the angle-resolved MChE by the competition between the 'DMI-locked' chiral spin textures and Zeeman effect, depending on the strength of the magnetic field. Our study demonstrates a state-of-the-art control of the magneto-chiral anisotropy, which paves the road for developing angle-resolved writing or detecting spintronic devices.

## Results

### Non-centrosymmetric atomic layer superlattice

The non-centrosymmetric superlattice with an A-B-C-C atomic-layer stacking order, where the inversion symmetry is absent at any point in space, was grown using reflection high energy electron diffraction assisted pulsed laser deposition. Details of the growth are described in Supporting Information (SI). The superlattice is coherently strained to the (001) STO substrate. Synchrotron X-ray diffraction (XRD) reveals a nearly tetragonal structure with the RuO$_6$ octahedrons mainly rotating around the $c$ axis ($\gamma \approx 14.5°$), while rotations around $a$, $b$ axes are significantly suppressed ($\alpha = \beta \approx 1.8°$) (see SI Section S1). Furthermore, we perform the cross-sectional scanning transmission electron microscopy (STEM) measurement and the energy-dispersive X-ray spectroscopy (EDS) imaging to demonstrate the element-resolved atomic structure of the superlattice. **Figure 1**a shows the high-angle annular dark field (HAADF) image of the sample probed along the [110] zone axis, and **Figure 1**b,c show the EDS elemental maps with the single-atomic-column resolution. The A layer, with the Ru sites (purple) partially substituted by Ti ions (green), is defined as $SrRu_{1-x}Ti_xO_3$ (SRTO). The B layer with negligible Ti substitution of Ru is approximated as SRO. To identify the elemental distributions quantitatively, both the Cliff-Lorimer method[26] and the Gaussian fitting method were applied (SI Section S1), which reveal that the A layer has about $x$=10~20% Ti substitution of the Ru ions (**Figure 1**d). Therefore, the A-B-C-C type SRTO-SRO-STO-STO atomic layer superlattice has been deliberately engineered in order to eliminate the inversion symmetry at the single-atomic level.

**The angle-resolved magneto-chiral anisotropy**

In order to probe the eMChE of the non-centrosymmetric superlattice with symmetry-allowed chiral DMI, we investigate MA under a series of static magnetic fields (***H***) through AMR. It is expected that the competition between the collinear Zeeman interaction ($\boldsymbol{H}\cdot\overline{\boldsymbol{S}}$) and the non-collinear DMI ($\boldsymbol{D}_{ij}\cdot(\boldsymbol{S}_i\times\boldsymbol{S}_j)$) leads to systematically varied MA. The experimental set-up is schematically shown in **Figure 1**e. The superlattice sample is fabricated into a Hall bar which is rotated under the static magnetic field in both right-handed (RH) and left-handed (LH) manners. The handedness of the rotation is used to stimulate the chiral response of MA.

Indeed, we observe the chiral response of MA under all applied magnetic fields, ranging from 9 T to nearly 0 T. Polar plots of AMR data measured under three characteristic fields, $\mu_0 H$ = 5, 3, and 2 T at 2 K, with sample rotated in both RH and LH manners, are shown in **Figure 1**f. The AMR is defined as $(\rho_{xx}(\theta) - \rho_{xx}^{max})/\rho_{xx}^{max}$, with $\rho_{xx}^{max}$ as the maximal resistance measured during the entire rotation. The AMR data measured at all the above fields do not present the PMA owned by the bulk SRO film. Instead, a systematically rotated uniaxial-MA (UMA) is observed, with the easy and hard axes rotating in the same handedness following the sample rotation handedness. Moreover, the magnitude of the UMA tilt depends on the strength of the field. For example, the 5 T field leads to a smaller UMA tilt, while the 2 T and 3 T fields induce larger UMA tilts. In addition, AMR measurements are also performed at 25 K and 50 K and similar results are observed (**Figure S4-S5**). Thus the sample rotation handedness and the strength of the external magnetic field together induce the angle and rotation-sense dependent UMA manipulation. These findings are all beyond the current understanding of conventional achiral magnetic materials where the magneto-crystalline anisotropy determines a fixed MA.

We study the systematic dependence of the UMA tilt on the magnetic field strength. **Figure 2**a shows the UMA revolution under sequentially tuned field from 5 T, 4 T, 3 T, etc., all the way to −9 T using the RH sample rotation. It is observed that the UMA tilting handedness is independent on the field direction, i.e., the positive and negative field induce the identical UMA tilting handedness. Only the sample rotation handedness matters. In **Figure S6**, The corresponding data of the LH rotation are shown and the same conclusion holds. The UMA tilting size is non-monotonically dependent on the field, with a maximum observed at $|\mu_0 H|$ = 2 T. Moreover, under $|\mu_0 H|$ = 2 T, the UMA is significantly distorted and there appear two non-overlapping directions of easy axes, with one near 30°/210° and the other at 225°/45° for 2/−2 T. These observations demonstrate the systematic, yet very sensitive, angular responses of the chiral UMA tuned by the external magnetic field.

To summarize the above observations, we show the revolution of the angle-resolved chiral UMA in **Figure 2**b,c. The AMR data is decomposed into three parts, as schematically shown in **Figure 2**b. The orange curve of '8' shape is a typical UMA behavior, which indicates the existence of reversible spins as the sample rotates from the positive to the negative direction of the easy axis. The green curve of the 'heart' shape originates from the non-reversible but rotatable spins, also from the UMA. The combination of the two explains most of the data in **Figure 2**a except those near $|\mu_0 H| = 2$ T. These two parts exhibit the same tilting angle of the chiral UMA, named as $\theta_\mathrm{p}$. Under the characteristic field of 2 T, the half '8' lobe of the reversible UMA rotates to a larger angle $\theta_q$, indicated by the purple half '8' lobe, when the sample rotates to the negative direction of the easy axis. In **Figure 2**c, the values of $\theta_\mathrm{p}$ -180° and $\theta_q$ -180° are shown as functions of the external field. $\theta_\mathrm{p}$ peaks around $|\mu_0 H| = 2$ T, while $\theta_\mathrm{q}$ is only observable at this field. **Figure S7** shows nice fittings of the AMR data based on the above model. The field dependence of $\theta_\mathrm{p}$ underlines the strong competition between the Zeeman energy and the chiral DMI energy, while the emergence of $\theta_\mathrm{q}$ at 2 T indicates a transition to an additional chiral state.

We perform the Hall measurement at the identical temperature of 2 K as the above AMR experiments. The result is shown in **Figure 2**d where the linear Hall contribution has been subtracted. We observe a strong non-linear peak feature as large as 1.7 μΩ·cm at the characteristic field of $|\mu_0 H| = 2$ T. The peak is gigantic as compared to the saturation part of 0.2 μΩ·cm. Evidently, the non-linear Hall data deviate from the typical anomalous Hall resistivity of an achiral magnet, which is expected to exhibit the same shape as the hysteresis loop (*M* vs *H*). However, the coercive field is observed to be about 1 T, which agrees well to the transition from the reversible spin dominated AMR above 1 T to the non-reversible spin dominated AMR below 1 T.

**MFM experiments**

In order to understand the above eMChE, we perform MFM measurements to probe the magnetic texture in real space. The superlattice was first magnetized at a perpendicular field of −7 T and then the field was gradually ramped to 0 T. A persistent stripe-domain contrast was observed from −7 T to −0 T, which corresponds to two-fold FFT peaks at $\pm \boldsymbol{q}_0 = [1.6~\mu m^{-1}, 2~\mu m^{-1}, 0]$ (**Figure 3**a-d). The observed robust stripe domains at fields as large as 7 T provide the existing ground of chiral spin textures near the domain walls, which is in line with the chiral tilt of UMA under the giant field of 9 T.

As the field reverses the sign and further increases from +0 T to +5 T, the domain contrast experiences dramatic changes (**Figure 3**d-i). At 0.5 T, the amplitude of the MFM domain contrast, which can be represented by the standard deviation of the MFM signals $(\delta f)_{\text{rms}}$, reaches a maximum indicating the strongest contrast between spin up and down regions, (**Figure 3**m), corresponding to the coercive field $H_c$. Note the slight difference to the 1 T coercive field in **Figure 2** is due to the temperature difference, i.e. 2 K for AMR versus 5 K for MFM . From 0 T to 1.4 T, the FFT peak intensity at $\boldsymbol{q}_0$ representing the stiffness of the stripe phase slightly decreases (**Figure 3**l). From 1.4 T to 1.8 T, the bright stripes break into small bubbles and the FFT peak intensity is drastically reduced. The smallest visible bubbles as being pointed out in **Figure 3**g have the sizes of ~100 nm, reaching the MFM lateral resolution. From 1.8 T to 2.4 T, the stripe phase is recovered, indicating the completed spin flipping. Simultaneously, the domain contrast drops dramatically (**Figure 3**m). The superlattice ultimately saturates at $H_s$ (2.4 T). Therefore, the revolution of the spin texture from stripe phase (−7 T to +1.4 T), to bubble phase (+1.4 T to +2.4 T), and back to stripe phase (+2.4 T to +5 T), strikingly coincides well to the AMR, which correlates the field-controlled spin textures characterized by MFM to the eMChE revealed by AMR. The stiff stripe phase corresponds to the robust $\theta_p$, while the bubble phase corresponds to the emergence of $\theta_p$ at the same critical field. Similar results are also found at 12 K (SI Section S4).

**Figure 3**j,k show $\rho_{xy}^{NL}$ and $\rho_{xx}$ measured at the same temperature with sample initially saturated in a large negative field. The saturation field of 2.4 T is confirmed by the saturation of $\rho_{xy}^{NL}$ shown in **Figure 3**j. While the coercive field of 0.5 T is confirmed by the $\rho_{xx}$ maximum shown in **Figure 3**k. The peak of $\rho_{xy}^{NL}$ is found to exist exactly at the critical field of 1.8 T ($H^*$), where the stripe phase completely disappears, the FFT peak intensity at $q_0$ is maximally suppressed and the inhomogeneous bubble phase emerges. Thus all the electrical measurements reflect well the spin texture transition.

**Atomistic simulation of rotating samples under external field**

We further perform atomistic simulations to study the magnetic structure evolution when the sample rotates under external field. The details of the simulation are shown in SI Section S5. In order to observe skyrmion bubble phase in the simulation, we obtained the required external field *H* with systematically varied DMI energy *D* (**Figure S12**).

The simulations were performed with the sample rotating in the RH manner, using a combination of $\mu_0 H$ = 3.5 T and $D$ = 2.3×10$^{-22}$ J. The results at an interval of 30° are shown in **Figure 4**a. Before the rotation, the spins are set to be fully polarized and the field is parallel to the sample normal ($\theta$ = 0°). Due to the large vertical component of ***H*** (***H***$_z$), the magnetization remains saturated when $\theta$ = 30° and 60°. It then develops into chiral Néel domain walls (see **Figure S13-S14**) when the Zeeman term ***H***$_z$·$\overline{S}$ is less or comparable to *D*, e.g., $\theta$ = 90° and 120°. At larger angles of $\theta$ = 180° and 210°, these chiral domain walls transform into skyrmion bubbles, which vary in the number and size during the sample rotation. When $\theta$ increases to 240°, bubbles become elongated along the in-plane field direction. When $\theta$ further increases from 270° to 330°, bubbles collapse into chiral domain walls and the magnetization becomes nearly saturated. **Figure 4**b summarizes the skyrmion bubble existing regime, which occupies an angular range significantly off-centered from 180°. The center of the range defined as $\theta_{Skr}$. Similar simulations are performed at other field values. A slightly larger but also off-centered skyrmion existing regime is found for $\mu_0 H$ = 4 T (**Figure 4**c). The size of $\theta_{Skr}$ as a function of $\mu_0 H$ is summarized in **Figure 4**d. At small ($\leq$ 3 T) or large $H$ ($\geq$ 5 T) regime, no skyrmion exists in the sample. In the intermediate region from 3.5 T to 4 T where skyrmions exist, $\theta_{Skr}$ slightly varies as a function of $H$. These findings explain the emergence of the large UMA tilting angle $\theta_q$ observed in the intermediate field regime.

The regime of the stripe phase is relatively harder to be defined. Nevertheless, **Figure 4**a indicates that the stripe phase appears mainly in 90°-150° and 270°-330° regimes, which are also off-centered from the symmetric angles of 90°/270°. Such observation holds true for smaller-field simulations absent of skyrmions (see **Figure S13**a for the $\mu_0 H = 3$ T simulation). In comparison, when the external field is strong enough that the Zeeman term $\boldsymbol{H}_z \cdot \overline{\boldsymbol{S}}$ dominates over $D$, the stripe phase is only observed at $\theta = 90°$ and 270° (see **Figure S13**b for the $\mu_0 H = 8$ T simulation). There are none $\theta$ off-centered (0°, 90°, 180°, 270°) magnetic textures at all, which corresponds to a typical achiral magnet with PMA. In addition to the simulation, we also performed density functional theory calculations, which reveal the robust ferromagnetic ground state with slight in-plane spin canting forming a non-coplanar spin texture (see SI Section S6).

**Discussion**

We propose that DMI governs the chiral UMA tilt, termed the 'DMI lock' effect. As schematically illustrated in top and bottom panels of **Figure 5**a,b, the DMI 'locks' chiral spin textures with $\boldsymbol{D}$ pointing out of the paper. In **Figure 5**a, when the sample rotates RH from $\theta = 0°$ toward 90°, the chiral spin texture shown in the top panel dominates over the one in the bottom panel, and becomes elongated in the spin right-pointing part (purple) and shortened in the left-pointing part (gray) due to the Zeeman energy gain. The overall magnetic moment is schematically shown in the top-right of **Figure 5**a middle panel, with $M_S$ pointing up and $M_H$ pointing right. When $\theta$ increases across 90°, a vertical-field inversed $M_S$ with a $M_H$ remained right-pointing by the parallel field, is not favored due to the increased DMI energy, as indicated by the dashed $M_S$ and $M_H$ at the bottom-right of **Figure 5**a middle panel. When $\theta$ increases across 180°, the $M_S$ and $M_H$ inversions are both favored by DMI and Zeeman energy terms, as shown in the bottom-left of **Figure 5**a middle panel. The bottom panel of **Figure 5**a shows the dominating DMI 'locked' spin texture for 180° < $\theta$ < 270°. When $\theta$ increases from 270° to 360°, similar conclusions hold as 90° < $\theta$ < 180°. **Figure 5**b shows the corresponding 'DMI lock' effect when the sample rotates left-handedly. Therefore, the chiral UMA tilts, as indicated by the light blue shades in **Figure 5**a,b middle panels, are schematically explained by the 'DMI lock' effect. The tilt angle depends on the $M_S/M_H$ ratio, which is controlled by the chiral spin texture (Néel stripe, skyrmion, etc.) revolution.

In comparison, **Figure 5**c shows the DMI 'unlocked' out-of-plane and in-plane magnetization components when an achiral magnet rotates under a static field. The top and bottom panels show the coexisting spin textures with opposite signs of helicity, due to the zero DMI. When the sample rotates, e.g. RH in the middle panel of **Figure 5**c, 4-fold symmetry of the $M_S + M_H$ is allowed due to the dominance of magnetocrystalline anisotropy (here presented as PMA) and the Zeeman energy. Therefore a chiral tilt of MA is not realizable in an achiral magnet.

**Conclusion**

In conclusion, we demonstrate an angle-resolved MChE in a non-centrosymmetric A-B-C-C (SRTO-SRO-STO-STO) atomic layer superlattice. The superlattice exhibits a chiral UMA tilt, depending on the sample rotation handedness. The tilt angle is systematically controlled by the static field $\mu_0 H$, ranging from 2º to 45º in the entire measured field range from 0 T to 9 T. Both MFM experiments and atomistic simulations reveal that chiral spin textures are closely correlated to the chiral tilts of UMA. The 'DMI lock' effect is employed to explain the angle-resolved MChE, which reflects the cooperation and competition between the Zeeman interaction and the DMI, with their delicate balance tipped by the rotation handedness and the field strength. Our finding of the angle-resolved MChE opens up a new horizon of employing atomic-level chiral spin interactions to manipulate spin orientations on the macroscopic scale, which shines light on developing novel angular-resolved sensing or writing types of chiral spintronic devices.

**Acknowledgements**

The authors thank Fuchun Zhang, Jiandi Zhang and Shilei Zhang for helpful discussions and Lingling Wang for helping EDS analysis. The work was financially supported by National Key R&D Program of China (Nos. 2022YFA1403000, 2022YFA1403300), the National Science Foundation of China (Nos. 52072244, 12104305, 12074080, 12274088, 11874054, 12104301), the Science and Technology Commission of Shanghai Municipality (Nos. 21JC1405000, 21PJ410800), Double First-Class Initiative Fund of ShanghaiTech University, the Startup Funds from the University of South Carolina and ShanghaiTech. The research used resources from Analytical Instrumentation Center (#SPST-AIC10112914) and CℏEM (EM02161943) in ShanghaiTech University, the Research Computing program under the Division of Information Technology at the University of South Carolina, and the Advanced Photon Source, a U.S. Department of Energy (DOE) Office of Science user facility at Argonne National Laboratory based on research supported by the U.S. DOE Office of Science-Basic Energy Sciences, under Contract No. DE-AC02-06CH11357.

**Contributions**

X.F.Z. designed the project and supervised the project with L.C., Z.Z., S.M. and W.W. Sample fabrication and the transport study were performed by J.Z., M.B. and L.C., with contributions from D.Q., J.L., F.Y., H.L. and G.C. The STEM-EDS study were done by Q.Y., L.C., J.-M.Z. Atomistic simulations were done by X.Z. and Z.Z. MFM measurements were performed by W.W., Q.L., X.D.Z., J.S. and W.W. analyzed the data. S.M. conducted the DFT calculations. H.C., H.Z. performed the synchrotron XRD study, with assistance from G.L. and Q.W. All authors contributed to the manuscript writing.

**Competing interests:** The authors declare that they have no competing interests.

**Data and materials availability:** All data needed to evaluate the conclusions in the paper are present in the paper and/or the Supplementary Materials. Additional data related to this paper may be requested from the authors.

**Figures**

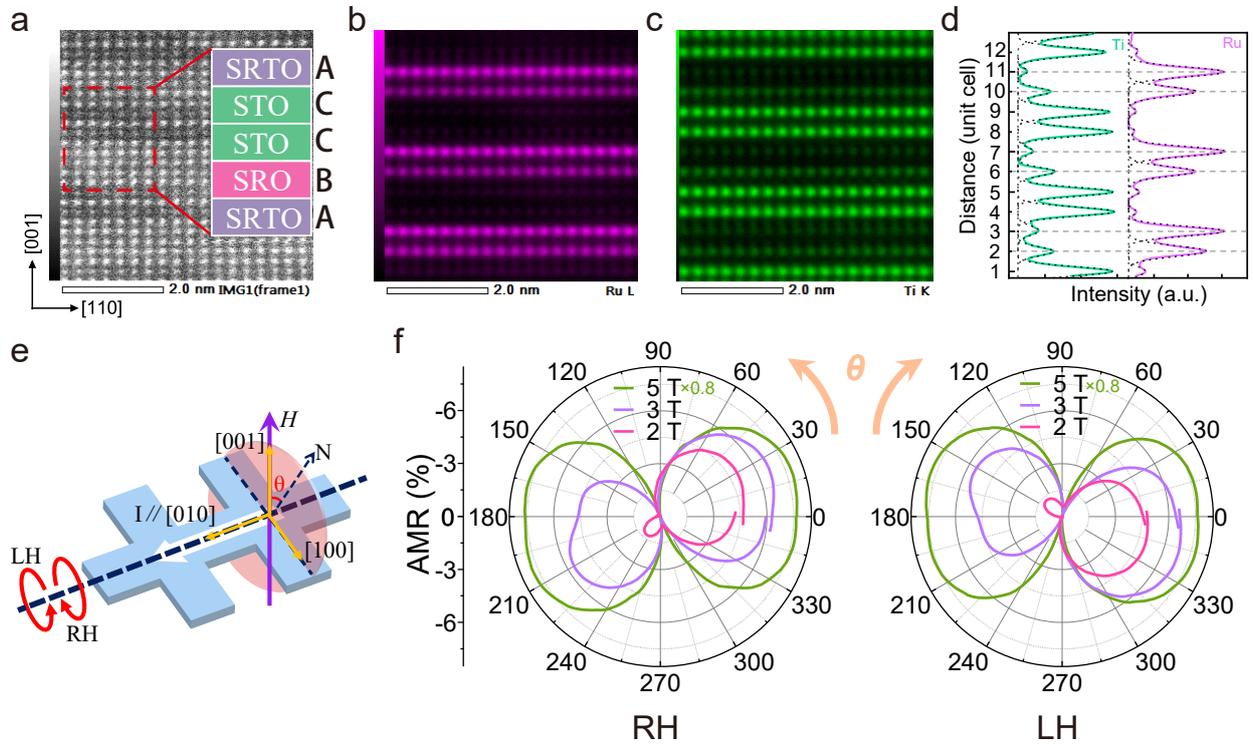

**Figure 1.** STEM, EDS, AMR measurements and atomistic simulations. (a) The cross-sectional STEM image. EDS mappings of (b) Ru and (c) Ti. (d) The Ru and Ti composition in each layer deduced from EDS. The direction of distance is indicated by the blue dash line in (a). (e) The geometry of the AMR measurements. The sample rotates around the current-flowing [010] direction. $\theta$ is the angle between the magnetic field and the surface normal direction. (f) Polar plots of the AMR measured under 5, 3, and 2 T at 2 K with the sample rotating in the RH and LH manners.

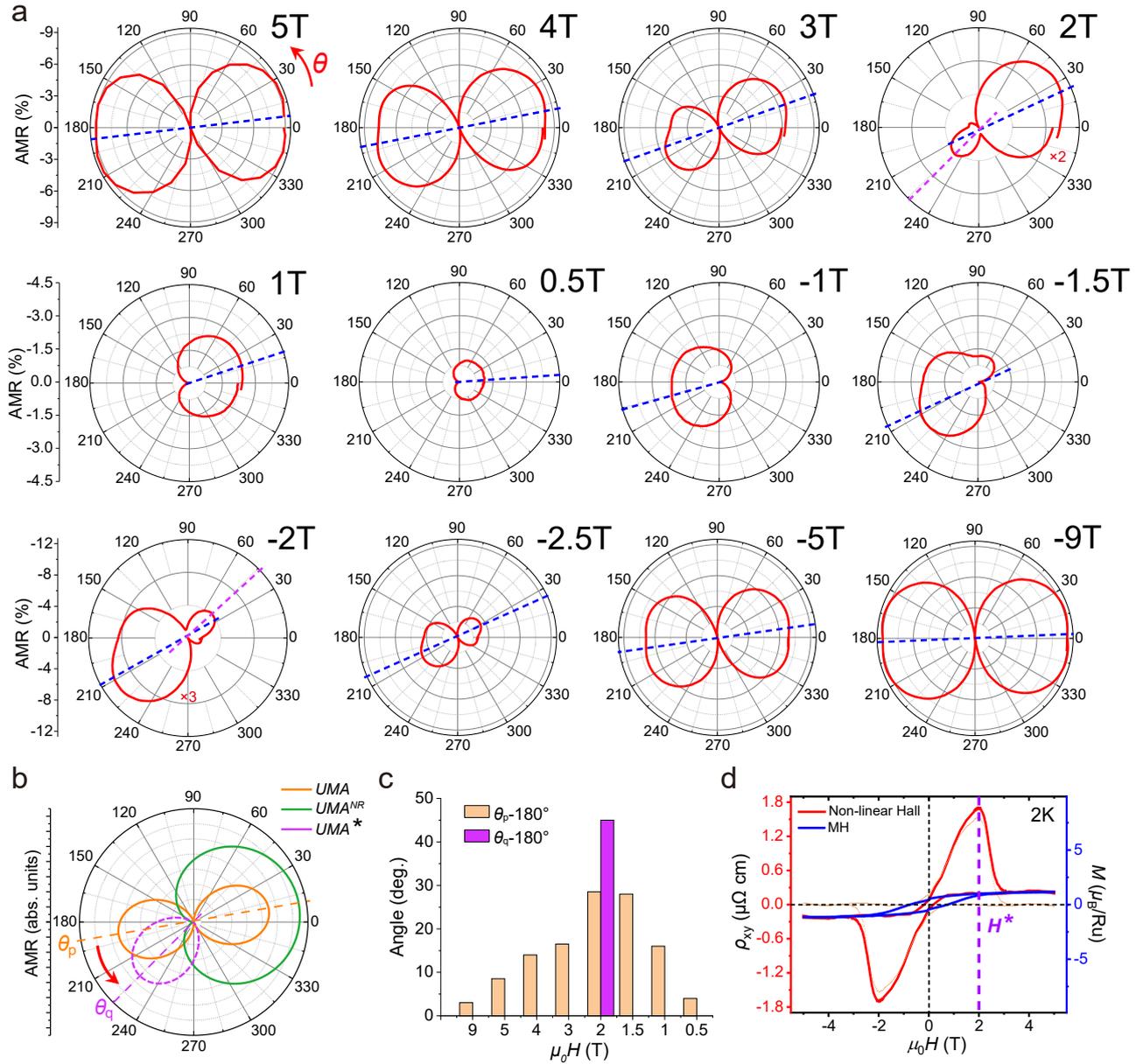

**Figure 2.** AMR and Hall measurements. (a) Polar plots of AMR measured continuously from 5 T to −9 T at 2 K with the RH sample rotation. (b) Schematic demonstration of the UMA (spin reversible), UMA$^{NR}$ (spin non-reversible) and UMA* components to the AMR. The orange dash line indicates the chiral rotation of the original UMA easy axis, and the purple dash line indicates the chiral rotation of the emergent UMA* easy axis. (c) Chiral rotation angles of the UMA easy axis ($\theta_p$) and the UMA* easy axis ($\theta_q$). (d) Non-linear Hall and the hysteresis loop (*M* vs *H*) measurement performed at 2 K.

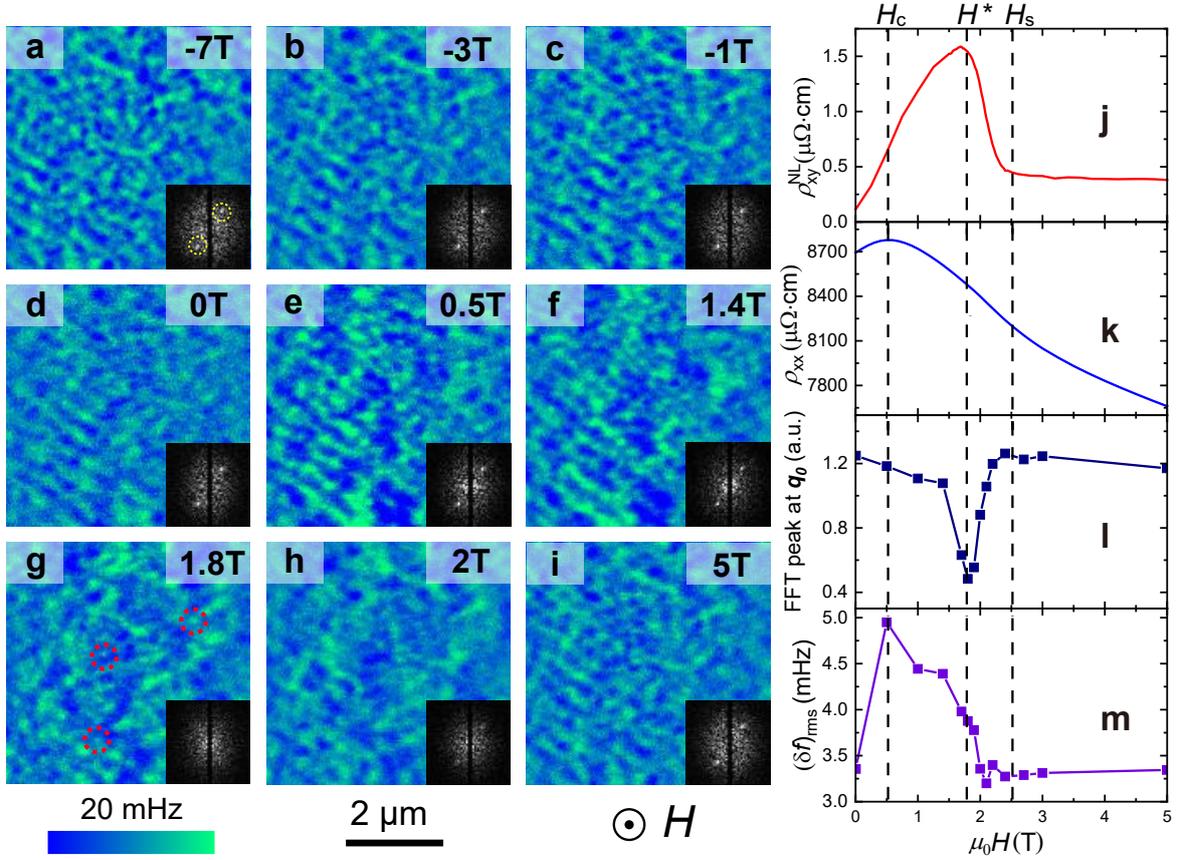

**Figure 3.** MFM performed at 5 K under sequentially increasing magnetic fields. (a-i) MFM images measured at magnetic fields from -7 T to 5 T. The insets show the FFT maps of the MFM images. Two-fold $\boldsymbol{q_0} = [1.6\ \mu m^{-1}, 2\ \mu m^{-1}, 0]$ FFT peaks in (a) were highlighted by yellow dashed circles. The magnetic bubbles are indicated by the red dashed circles in (g) and the size is estimated to be ~100 nm. (j-m) The $\rho_{xy}^{NL}$, $\rho_{xx}$, the FFT peak at $\boldsymbol{q}_0$ and standard deviation of MFM signals $(\delta f)_{rms}$. The dashed lines indicate the three critical fields, $H_c$, $H^*$, and $H_s$.

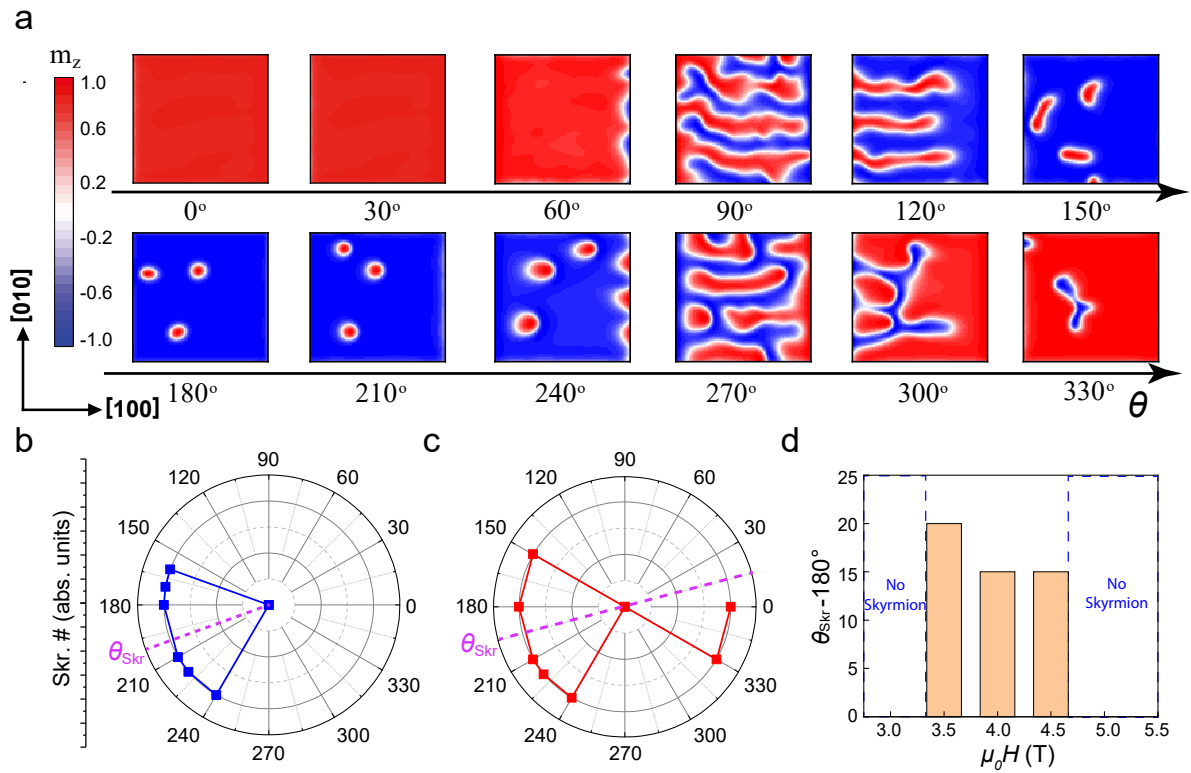

**Figure 4.** The simulated magnetic texture under rotating field. (a) Evolution of magnetic texture at 3.5 T. The initial magnetization profile at 0° is produced at 8 T and then $H_{ext}$ is reduced to 3.5 T for the rotating process. (b) Field region where skyrmion exists at $H_{ext}$ = 3.5 T and (c) $H_{ext}$ = 4 T. The dashed line denotes the middle of the skyrmion region. (d) The angles corresponding to the middle of the skyrmion region.

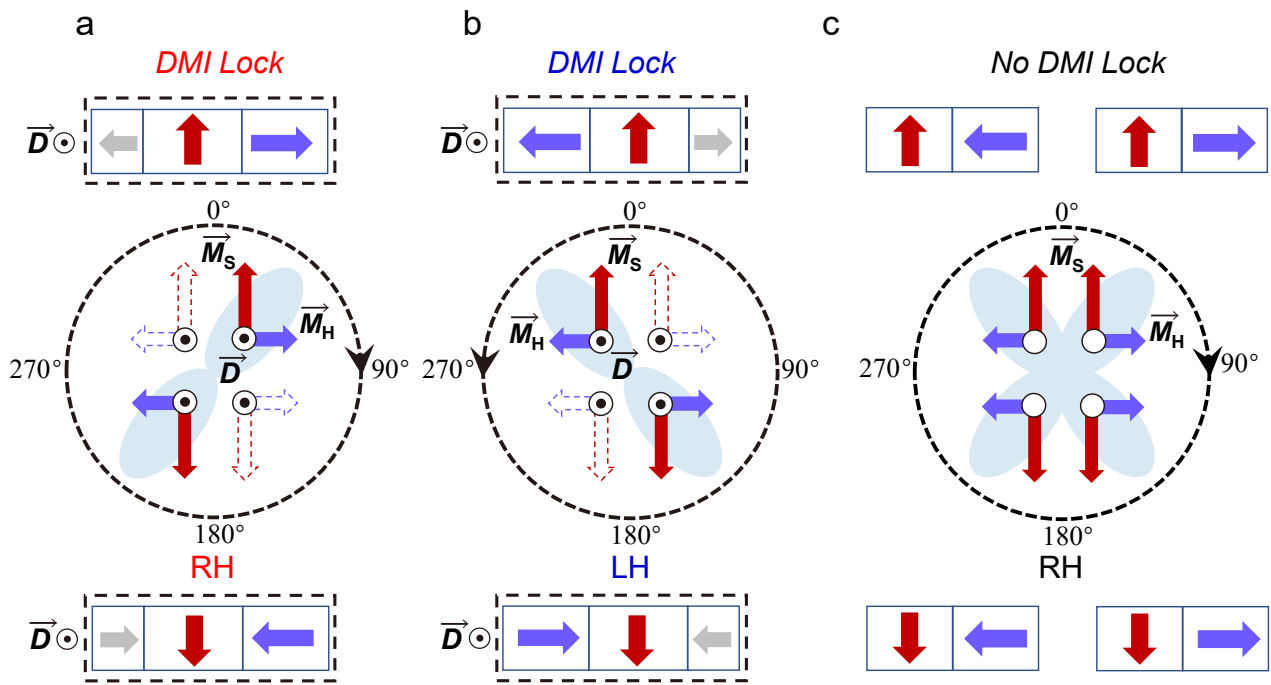

**Figure 5.** Illustrations of the chiral UMA tilts under DMI Lock with the sample rotating (a) right-handed and (b) left-handed. (c) The four-fold MA of a sample without DMI Lock under right-handed rotation.